\documentclass[reprint,aps,nofootinbib,superscriptaddress,longbibliography]{revtex4-1}
\usepackage[latin9]{inputenc}
\usepackage{color}
\usepackage{array}
\usepackage{multirow}
\usepackage{amsmath}
\usepackage{amssymb}
\usepackage{graphicx}
\usepackage{wasysym}
\usepackage[unicode=true,pdfusetitle,
 bookmarks=true,bookmarksnumbered=false,bookmarksopen=false,
 breaklinks=false,pdfborder={0 0 1},backref=false,colorlinks=true]
 {hyperref}
\hypersetup{
 pdfborderstyle=,pdfborderstyle={},pdfborderstyle={},pdfborderstyle={},pdfborderstyle={}}

\makeatletter

\providecommand{\tabularnewline}{\\}

\usepackage{bm}
\usepackage{fancyhdr}
\usepackage{amsfonts}
\usepackage[mathscr]{eucal}
\usepackage{soul}
\usepackage{psfrag}
\usepackage{etoolbox}

\newcommand{\ass}{\stackrel{\textup{\tiny def}}{=}}

\newcommand{\cv}[1]{\Delta_{#1}}

\newtheorem{theorem}{Theorem}
\newtheorem{lemma}{Lemma}

\makeatother

\begin{document}
\global\long\def\sgn{\mathrm{sgn}}%
\global\long\def\ket#1{|#1\rangle}%
\global\long\def\bra#1{\langle#1|}%
\global\long\def\sp#1#2{\langle#1|#2\rangle}%
\global\long\def\abs#1{\left|#1\right|}%

\title{Bounds on nonlocal correlations in the presence of signaling and their
application to topological zero modes}
\author{Avishy Carmi}
\affiliation{Center for Quantum Information Science and Technology \& Faculty of
Engineering Sciences, Ben-Gurion University of the Negev, Beersheba
8410501, Israel}
\author{Yaroslav Herasymenko}
\affiliation{Instituut-Lorentz for Theoretical Physics, Leiden University, Leiden,
NL-2333 CA, The Netherlands}
\author{Eliahu Cohen}
\affiliation{Faculty of Engineering and the Institute of Nanotechnology and Advanced
Materials, Bar Ilan University, Ramat Gan 5290002, Israel}
\author{Kyrylo Snizhko}
\affiliation{Department of Condensed Matter Physics, Weizmann Institute of Science,
Rehovot, 76100 Israel}
\begin{abstract}
Bell's theorem renders quantum correlations distinct from those of
any local-realistic model. Although being stronger than classical
correlations, quantum correlations are limited by the Tsirelson bound.
This bound, however, applies for \emph{Hermitian}, \emph{commutative}
operators corresponding to \emph{non-signaling} observables in Alice's
and Bob's spacelike-separated labs. As an attempt to explore theories
beyond quantum mechanics and analyze the uniqueness of the latter,
we examine in this work the extent of non-local correlations when
relaxing these fundamental assumptions, which allows for theories
with non-local signaling. We prove that, somewhat surprisingly, the
Tsirelson bound in the Bell--CHSH scenario, and similarly other related
bounds on non-local correlations, remain effective as long as we maintain
the Hilbert space structure of the theory. Furthermore, in the case
of Hermitian observables we find novel relations between non-locality,
local correlations, and signaling.

We demonstrate that such non-local signaling theories are naturally
simulated by quantum systems of parafermionic zero modes. We numerically
study the derived bounds in parafermionic systems, confirming the
bounds' validity yet finding a drastic difference between correlations
of ``signaling'' and ``non-signaling'' sets of observables. We
also propose an experimental procedure for measuring the relevant
correlations.
\end{abstract}
\maketitle

\section{Introduction}

In a Bell test \citep{Bell1964,Brunner2014}, Alice and Bob measure
pairs of particles (possibly having a common source in their past)
and then communicate in order to calculate the correlations between
these measurements. The strength of empirical correlations enables
one to characterize the underlying theory. In quantum mechanics, the
above procedure corresponds to local measurements of Hermitian operators
$A_{0}$/$A_{1}$ on Alice's side and $B_{0}/B_{1}$ on Bob's side.
The correlators are defined using the quantum expectation value $c_{ij}=\langle A_{i}B_{j}\rangle$
and, when the operators have eigenvalues $\pm1$, it can be shown
that the CHSH parameter obeys 
\begin{equation}
|\mathcal{B}|\equiv|c_{00}+c_{10}+c_{01}-c_{11}|\le2\sqrt{2},\label{eq:Tsirelson_standard}
\end{equation}
which is known as the Tsirelson bound \citep{Cirelson1980}. Stronger
bounds on the correlators (i.e., bounds from which the Tsirelson bound
can be derived) were proposed, e.g., by Uffink \citep{Uffink2002}
and independently by Tsirelson, Landau and Masanes (TLM) \citep{Tsirelson1987,Landau1988,Masanes2003}.
The latter implies that 
\begin{equation}
|c_{00}c_{10}-c_{01}c_{11}|\le\sum_{j=0,1}\sqrt{(1-c_{0j}^{2})(1-c_{1j}^{2})}.\label{eq:TLM_standard}
\end{equation}
The TLM inequality is known to be necessary and sufficient for the
correlators $c_{ij}$ to be realizable in quantum mechanics \citep{Tsirelson1987,Landau1988,Masanes2003}
(implying, in particular, that if a set of correlators satisfies Eq.~(\ref{eq:TLM_standard}),
it necessarily satisfies Eq.~(\ref{eq:Tsirelson_standard}); the
converse is not true). Importantly, when calculating $\mathcal{B}$
in any local-realistic model it turns out that $\abs{\mathcal{B}}\le2$,
which is a famous variant of Bell's theorem known as the Clauser-Horne-Shimony-Holt
(CHSH) inequality \citep{Clauser1969}, which provides a measurable
distinction between correlations achievable in local-realistic models
and in quantum theory. These bounds, however, are not enough for fully
characterizing the Alice-Bob quantum correlations. For the latter
task, the Navascues-Pironio-Acin (NPA) hierarchical scheme of semidefinite
programs was proposed \citep{Navascues2008}.

All the above works plausibly assuming that Alice's and Bob's measurements
are described by spatially local and Hermitian operators, implying
that $[A_{i},B_{j}]=0$ for all $i,j$. As such, they cannot lead
to superluminal signaling between Alice and Bob.

Trying, on the one hand, to generalize some of the above results,
and on the other hand to pin-point the core reason they work so well,
we relax below these two assumptions and examine the consequences
of complex-valued correlations emerging from non-Hermitian non-commuting
Alice/Bob operators. We thus allow a restricted form of signaling
between the parties (similar to the one in \citep{Silman2013}), but
we maintain the Hilbert space structure, as well as other core ingredients
of quantum mechanics. Surprisingly, the Tsirelson bound and TLM inequality
remain valid in this generalized setting. Apart from that, we find
intriguing relations between nonlocality, local correlations of Alice
(or Bob), and signaling in the case of Hermitian yet non-commuting
observables.

Considering non-Hermitian non-commuting observables may seem far from
any sensible model. To alleviate this impression, we study an explicit
example of a parafermionic system, which is a proper quantum system
that provides a natural setting for comparing commuting and non-commuting
sets of observables. The natural observables in the parafermionic
system happen to be non-Hermitian. Parafermions (or rather parafermionic
zero modes) are topological zero modes that generalize the better-known
Majorana zero modes \citep{Alicea2012,Leijnse2012,Beenakker2013}.
Parafermions can be realized in various quasi-one-dimensional systems
\citep{Lindner2012,Cheng2012,Vaezi2013,Clarke2013,Mong2014,Klinovaja2014,Klinovaja2014a},
see \citep{Alicea2016} for a comprehensive review. Similarly to the
case of Majoranas, observables in a system of parafermions are inherently
non-local as they comprise at least two parafermionic operators hosted
at different spatial locations. In the case of Majoranas, this nonlocality
is known to have manifestations through the standard CHSH inequality
\citep{Romito2017}. We do not follow the investigation line of Ref.~\citep{Romito2017},
but rather investigate a different aspect of nonlocality, which is
absent for Majoranas yet present for parafermions.

Specifically, we construct two examples. In the first, the system
of parafermions is split into two spatially separated parts, $A$
and $B$, with commuting observables $[A_{i},B_{j}]=0$. In the second
example, Alice's and Bob's parts are still spatially separated; the
local permutation properties of $A_{0},A_{1}$, as well as those of
$B_{0},B_{1}$ are exactly the same as in the first example, yet $[A_{i},B_{j}]\neq0$.
This property alone has the potential to contradict relativistic causality
since we have spatially separated observables which do not commute
and thus allow for superluminal signaling (thus these systems can
indeed simulate the case of non-Hermitian signaling operators). However,
as we explain in Sec.~\ref{sec:parafermions_Alice's-and-Bob's_observables},
in order to measure their respective observables, Alice and Bob in
our system must share a common region of space, which resolves the
paradox. In this sense, Alice and Bob can be thought of as two experimenters
acting on the same system. Therefore, the system of parafermions does
not constitute a system in which the spatial and quantum mechanical
notions of locality disagree. However, it simulates such a system
(with spatial locality interpreted in a very naive way). Using these
examples we investigate the theoretical bounds on correlations. We
find that both systems obey the derived bounds. However, the maximal
achievable correlations in the truly local system (first example)
are significantly weaker than those of the non-local one.

Before we present our results in the next sections, one comment is
due. One may think that investigating Bell-CHSH correlations with
$[A_{i},B_{j}]\neq0$ is an abuse of notation. Originally introduced
for distinguishing local-realistic theories from the standard quantum
theory, the Bell-CHSH inequalities imply the use of conditional probabilities
$P(a,b|i,j)$ that are defined in both. With $[A_{i},B_{j}]\neq0$,
the correlators that have the same operator form are expressed not
through probability distributions $P(a,b|i,j)$ but rather through
quasiprobability distributions $W(a,b|i,j)$, cf.~Appendix~\ref{appendix:measuring_correlations_with_weak_measurements}.
Therefore, a formal replacement of commuting operators with non-commuting
ones may seem an illegitimate operation in this context. We would
like to emphasize that the key to comparing properties of different
theories is considering objects that are defined in these theories
in an operationally identical way. This is the reason that local-realistic
theories are compared to quantum mechanics not in terms of the joint
probability distribution $P(a_{0},a_{1},b_{0},b_{1})$ (that does
not exist in quantum theory when $[A_{i},A_{j}]\neq0$ and/or $[B_{i},B_{j}]\neq0$)
but in terms of $P(a,b|i,j)$ conditioned on the choice of observables:
$P(a,b|i,j)$ are defined in both theories and can be measured by
the same measurement procedure. Since our aim here is to compare the
standard quantum theory with that allowing for $[A_{i},B_{j}]\neq0$,
working in the language of correlators that are defined and can be
measured (even if they are complex) by means of weak measurements
in both theories \citep{Bulte2018} is a natural decision. We thus
compare nonlocal theories having a Hilbert space structure, rather
than a probabilistic structure (common, e.g., to local hidden variables
theories and quantum mechanics, but not to the post-quantum theories
discussed here). However, in the case of the standard quantum theory,
the correlation functions (and thus our new bounds) can be expressed
in terms of $P(a,b|i,j)$, making them new bounds on the possible
probability distributions in the standard quantum theory.

In what follows, we start in Sec. \ref{sec:1} by defining an operator-based
(rather than probability-based) notion of complex correlations arising
in nonlocal, non-Hermitian systems admitting signaling and then find
the generalized inequalities bounding them. Importantly, this notion
has an operational sense in terms of weak measurements, as discussed
in the Appendix \ref{sec:measuring_non-commuting_observables}. In
Sec. \ref{sec:2}, we review parafermionic systems and show they can
simulate such non-Hermitian signaing systems. We then numerically
prove they are indeed bounded by the proposed bounds. Sec. \ref{sec:3}
concludes the paper. Some technical details appear in the Appendices.

\section{Analytic results for correlations of general non-Hermitian non-commuting
operators}

\label{sec:1}

Below we prove a number of bounds on quantum correlations of non-Hermitian
non-commuting operators. We generalize the Tsirelson and the TLM bounds
(Theorems \ref{Theorem 1} and \ref{Theorem 2}, which have been previously
derived for Hermitian commuting operators, see \citep{Carmi2018a})
and derive previously unknown bounds (Theorems \ref{Theorem 3} and
\ref{Theorem 4}, which are applicable to the Hermitian, non-signaling
case as well). Here we introduce the bounds and discuss them, while
their proofs are deferred to Sec.~\ref{subsec:Proofs-of-analytic-bounds}.
The bounds are expressed in terms of Pearson correlation functions
of operators $X$ and $Y$ defined as
\begin{equation}
C(X,Y)=\frac{\langle XY^{\dagger}\rangle-\langle X\rangle\langle Y\rangle^{\dagger}}{\cv{X}\cv{Y}},\label{eq:C(X,Y)_def}
\end{equation}
where $\cv{X}=\sqrt{\langle XX^{\dagger}\rangle-\abs{\langle X\rangle}^{2}}$
is the variance of $X$ (which is assumed to be non-zero), and averaging
is performed with respect to some state $|\psi\rangle$ in the Hilbert
space. This definition is a straightforward generalization of the
usual Pearson correlation between commuting Hermitian operators. The
Pearson correlations reduce to the standard $c_{XY}=\langle XY\rangle$
for Hermitian $X$ and $Y$ on states $\ket{\psi}$ such that $\langle X\rangle=\langle Y\rangle=0$
and $\cv{X}=\cv{Y}=1$. We note that $C(X,Y)$ is ill-defined when
$\cv{X}=0$ or $\cv{Y}=0$; yet, as we show in Sec.~\ref{subsec:Proofs-of-analytic-bounds},
$\abs{C(X,Y)}\leq1$ everywhere, including the vicinity of such special
points.

For the case of commuting operators $X$, $Y$, the definition of
$C(X,Y)$ can be expressed in terms of the joint probability distributions,
and our below bounds can be thought of as restricting the possible
probability distributions in quantum theory. When $X$ and $Y$ do
not commute, this is not the case, which defies the notions that conventionally
underlie Bell inequalities. However, our aim here is not to analyze
complex local hidden variables models but rather to examine general
models which are manifestly nonlocal. In particular, we wish to analyze
whether known bounds on quantum correlations remain effective when
generalized to cases of non-Hermitian signaling operators. We argue
that these complex correlations are physically meaningful because
there is an empirical protocol for measuring them. That operational
meaning of the above correlations in terms of weak measurements is
given in Appendix~\ref{sec:measuring_non-commuting_observables}.
Alternatively, for the case of non-commuting observables, $C(X,Y)$
can be expressed in terms of quasiprobability distibutions, and thus
our bounds restrict possible quasiprobability distributions in that
case. We discuss this in detail in Appendix~\ref{sec:complex_probabilities}.

We now discuss the bounds on Alice-Bob correlations.

\begin{theorem} \textbf{(Generalized Tsirelson bound)}. Define $\mathcal{B}\ass C(A_{0},B_{0})+C(A_{1},B_{0})+C(A_{0},B_{1})-C(A_{1},B_{1})$
as the complex-valued Bell-CHSH parameter of any operators $A_{i}$
and $B_{j}$. The following holds 
\begin{multline}
\abs{\mathcal{B}}=\sqrt{\mathrm{Re}(\mathcal{B})^{2}+\mathrm{Im}(\mathcal{B})^{2}}\\
\leq\sqrt{2}\left[\sqrt{1+\mathrm{Re}(\eta)}+\sqrt{1-\mathrm{Re}(\eta)}\right]\\
\leq2\sqrt{2},\label{Theorem 1 ineq}
\end{multline}
where $\eta$ is either $C(A_{0},A_{1})$ or $C(B_{0},B_{1})$ (the
one having the larger $\abs{\mathrm{Re}(\eta)}$ among them will give
rise to a tighter inequality). \label{Theorem 1} \end{theorem}

Despite the fact that $C(X,Y)\neq c_{XY}$, the Bell-CHSH parameter
defined through $C(X,Y)$ obeys the same Tsirelson bound as for $c_{XY}$
in Eq.~(\ref{eq:Tsirelson_standard}). Moreover, the proof of the
Tsirelson bound for $C(X,Y)$ is valid independently of whether $[A_{i},B_{j}]=0$.
A somewhat tighter bound (the middle row of Eq.~(\ref{Theorem 1 ineq}))
is obtained in terms of $\eta$ that expresses on-site correlations
on Alice's or Bob's side. This is also insensitive to whether $[A_{i},B_{j}]=0$.

\begin{theorem} \textbf{(Generalized TLM bound)}. The following holds
for any operators $A_{i}$, $B_{j}$, $i,j\in\{0,1\}$, 
\begin{multline}
\abs{C(B_{0},A_{0})^{\dagger}C(B_{0},A_{1})-C(B_{1},A_{0})^{\dagger}C(B_{1},A_{1})}\\
\leq\sum_{j=0,1}\sqrt{(1-\abs{C(B_{j},A_{0})}^{2})(1-\abs{C(B_{j},A_{1})}^{2})}.\label{Theorem 2 ineq}
\end{multline}
 \label{Theorem 2} \end{theorem}

Similarly to the previous theorem, this bound is insensitive to whether
$[A_{i},B_{j}]=0$ and has the same form as the standard TLM bound,
Eq.~(\ref{eq:TLM_standard}), modulo replacing $C(X,Y)$ with real-valued
$c_{XY}$.

We note in passing that our bounds apply to both operators with bounded
and unbounded spectrum. Implementing Bell tests in mesoscopic systems
often requires dealing with operators having an unbounded spectrum,
cf.~Ref.~\citep{Bednorz2011}. Our theorems~\ref{Theorem 1} and
\ref{Theorem 2} may thus be useful for studies in such systems.

\begin{theorem} \textbf{(Relation between nonlocality, local correlations,
and signaling)}. Let $\mathcal{B}$ be the complex-valued Bell-CHSH
parameter defined in Theorem \ref{Theorem 1}. Then, 
\begin{equation}
\left(\frac{\mathrm{Re}(\eta)}{2}\right)^{2}+\left(\frac{\mathrm{Re}(\mathcal{B})}{2\sqrt{2}}\right)^{2}+\left(\frac{\mathrm{Im}(\mathcal{B})}{2\sqrt{2}}\right)^{2}\leq1.\label{Theorem 3 ineq}
\end{equation}
\label{Theorem 3} \end{theorem}

\begin{figure}
\centering 

\includegraphics[width=1\columnwidth]{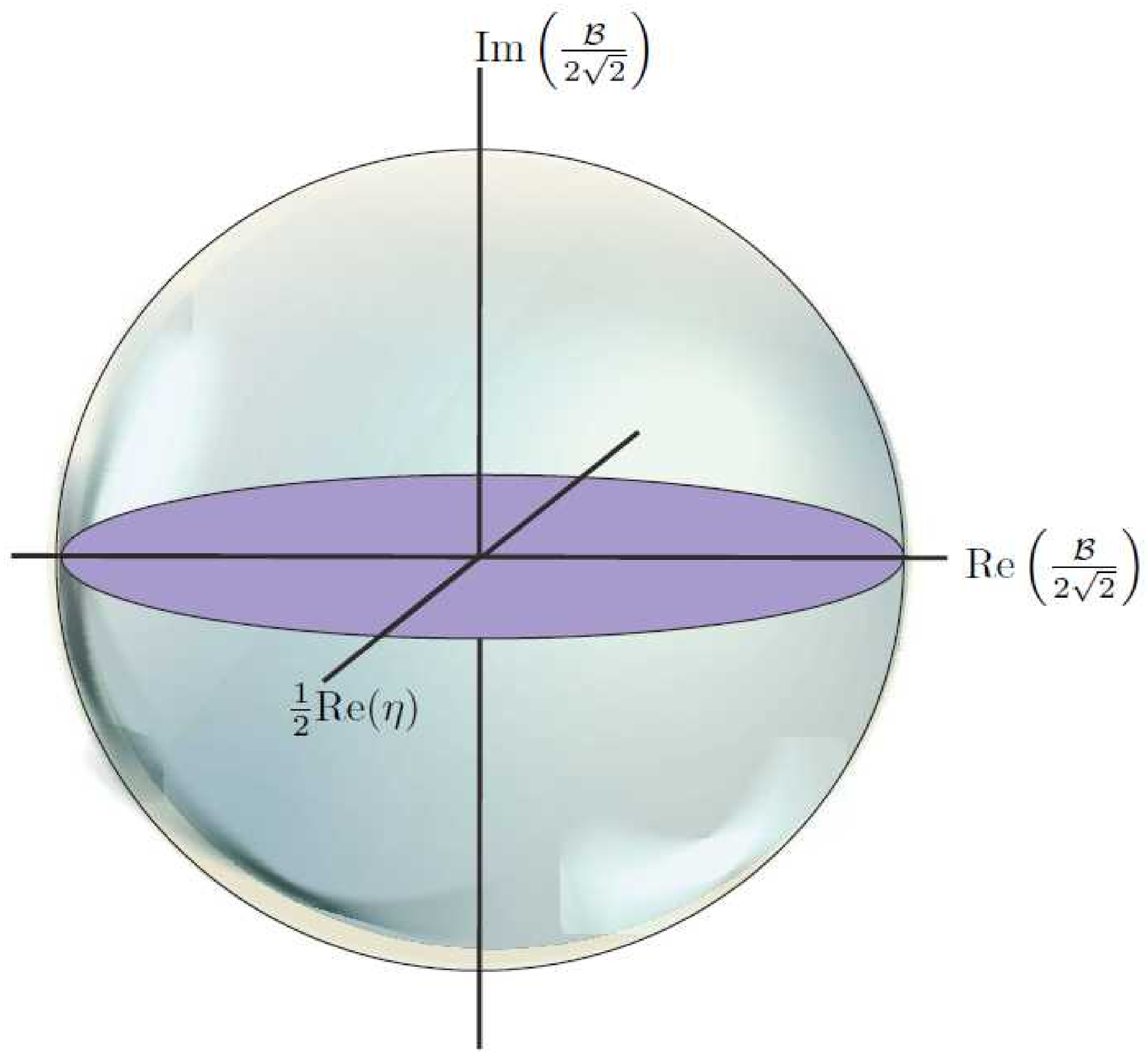} 



{\small{}{}{}{}{}{}\caption{{\small{}{}{}{}{}{}Local correlations \protect $\mathrm{Re}(\eta)/2$,
nonlocality \protect $\mathrm{Re}(\mathcal{B})/(2\sqrt{2})$, and
quantum signaling \protect $\mathrm{Im}(\mathcal{B})/(2\sqrt{2})$
are confined to the unit ball as implied by Theorem \ref{Theorem 3}.
The relation between the nonlocality and local correlations in an
ordinary quantum theory \protect {}with commutative Hermitian observables,
where the signaling parameter $\mathrm{Im}(\mathcal{B})=0$, is represented
by the purple section \protect {}within the ball. \protect}}
\label{fig:sphere} }{\small\par}

\end{figure}

This bound is also valid independently of $[A_{i},B_{j}]=0$. In the
case of Hermitian $A_{i}$, $B_{j}$ that obey $[A_{i},B_{j}]=0$,
$C(A_{i},B_{j})$ is real, implying $\mathrm{Im}(\mathcal{B})=0$.
If $A_{i}$ and $B_{j}$ are Hermitian but do not mutually commute,
there can appear imaginary components to $C(A_{i},B_{j})$ and $\mathcal{B}$.
Therefore, this relation may be interpreted as a constraint on non-local
correlations (represented by $\mathrm{Re}(\mathcal{B})/(2\sqrt{2})$),
local on-site correlations ($\mathrm{Re}(\eta)/2$), and signaling
(represented by $\mathrm{Im}(\mathcal{B})/(2\sqrt{2})\neq0$). These
three quantities are thus confined to the unit ball, see Figure~\ref{fig:sphere}.

\begin{theorem} \label{Theorem 4} Let $\mathcal{B}$ be the complex-valued
Bell-CHSH parameter defined in Theorem \ref{Theorem 1}. In the case
of isotropic correlations, $C(A_{i},B_{j})=(-1)^{ij}\varrho$ (such
that $\mathcal{B}=4\varrho$) for some complex-valued $\varrho$,
\begin{equation}
\abs{\eta}^{2}+\left(\frac{\mathrm{Re}(\mathcal{B})}{2\sqrt{2}}\right)^{2}+\left(\frac{\mathrm{Im}(\mathcal{B})}{2\sqrt{2}}\right)^{2}\leq1.\label{Theorem 4 ineq}
\end{equation}

\end{theorem}

Note that Eq.~(\ref{Theorem 4 ineq}) provides a tighter bound than
Eq.~(\ref{Theorem 3 ineq}). However, Eq.~(\ref{Theorem 4 ineq})
is proved under the rather restrictive assumption of $C(A_{i},B_{j})=(-1)^{ij}\varrho$.
This is a valid assumption within non-signaling theories in the following
sense. Reference~\citep{Masanes2006} argued that the standard Bell-CHSH
parameter $\mathcal{B}=c_{00}+c_{10}+c_{01}-c_{11}$ for $\pm1$-valued
observables in a non-signaling theory (not necessarily classical or
quantum) can always be maximized on a state satisfying $c_{ij}=(-1)^{ij}\rho$
with a real $\rho$. While the statement of Ref.~\citep{Masanes2006}
was proved for the standard correlations $c_{XY}$ (and not our $C(X,Y)$)
and maximizing the l.h.s. of Eq.~(\ref{Theorem 4 ineq}) is not equivalent
to maximizing $\abs{\mathcal{B}}$, one might hope that the possibility
of arranging $C(A_{i},B_{j})=(-1)^{ij}\varrho$ is related to non-signaling,
and the bound of Eq.~(\ref{Theorem 4 ineq}) would discriminate the
cases of $[A_{i},B_{j}]=0$ and $[A_{i},B_{j}]\neq0$. We provide
some numerical evidence for the last statement in Sec.~\ref{sec:2}.

\subsection{\label{subsec:Proofs-of-analytic-bounds}Proofs of analytic bounds}

\begin{lemma} \textbf{(Generalized uncertainty relations, see Ref.~\citep{Carmi2019}
for elaboration on the term)}. Denote by $X_{1},\ldots,X_{n}$, a
number of operators. Let $C$ be an $n\times n$ Hermitian matrix
whose $ij$-th entry is 
\begin{equation}
C(X_{i},X_{j})=\frac{\langle X_{i}X_{j}^{\dagger}\rangle-\langle X_{i}\rangle\langle X_{j}\rangle^{\dagger}}{\cv{X_{i}}\cv{X_{j}}},\label{eq:correlations_def}
\end{equation}
where $\cv{X}=\sqrt{\langle XX^{\dagger}\rangle-\abs{\langle X\rangle}^{2}}$
is the uncertainty in $X$ (which is assumed to be non-zero). Then
$C\succeq0$, i.e., it is positive semidefinite. \end{lemma}

\emph{Proof.} Denote $|\psi\rangle$, the underlying quantum state.
For any $n$-dimensional vector, $v^{T}=[v_{1},\ldots,v_{n}]$, it
follows that 
\begin{equation}
v^{T}DCD^{T}v=\langle\phi|\phi\rangle\geq0,
\end{equation}
where $D$ is a (positive semidefinite) diagonal matrix whose entries
are $D_{ii}=\cv{X_{i}}$, and $|\phi\rangle=\sum_{i=1}^{n}v_{i}\left(X_{i}-\langle X_{i}\rangle\right)|\psi\rangle$.
Therefore, $DCD^{T}\succeq0$ and so is $C\succeq0$. $\Square$

Applying this lemma to two operators, $X_{1},X_{2}$, one obtains
that $\abs{C(X_{1},X_{2})}\leq1$, implying that the correlation functions
are bounded even near $\cv{X_{1,2}}=0$.

\textbf{Theorem \ref{Theorem 1}}.\emph{ Proof}. Construct the matrix
$C$ for the operators $A_{0}$, $A_{1}$, and $B_{j}$, 
\begin{multline}
\begin{bmatrix}C(B_{j},B_{j}) & C(B_{j},A_{1}) & C(B_{j},A_{0})\\
C(B_{j},A_{1})^{\dagger} & C(A_{1},A_{1}) & C(A_{0},A_{1})\\
C(B_{j},A_{0})^{\dagger} & C(A_{0},A_{1})^{\dagger} & C(A_{0},A_{0})
\end{bmatrix}\\
=\begin{bmatrix}1 & C(B_{j},A_{1}) & C(B_{j},A_{0})\\
C(B_{j},A_{1})^{\dagger} & 1 & \eta\\
C(B_{j},A_{0})^{\dagger} & \eta^{\dagger} & 1
\end{bmatrix}\succeq0,\label{semid}
\end{multline}
where $\eta\ass C(A_{0},A_{1})$. By the Schur complement condition
for positive semidefiniteness this is equivalent to 
\begin{equation}
C^{A}\ass\begin{bmatrix}1 & \eta\\
\eta^{\dagger} & 1
\end{bmatrix}\succeq\begin{bmatrix}C(B_{j},A_{1})^{\dagger}\\
C(B_{j},A_{0})^{\dagger}
\end{bmatrix}\begin{bmatrix}C(B_{j},A_{1}) & C(B_{j},A_{0})\end{bmatrix}.\label{eq:ineq1}
\end{equation}
Let $v_{j}^{T}=[(-1)^{j},\,1]$. The above inequality implies 
\begin{multline}
2(1+(-1)^{j}\mathrm{Re}(\eta))=v_{j}^{T}C^{A}v_{j}\\
\geq\abs{C(B_{j},A_{0})+(-1)^{j}C(B_{j},A_{1})}^{2}.
\end{multline}
This together with the triangle inequality yield 
\begin{multline}
\abs{\mathcal{B}}\leq\sum_{j=0,1}\abs{C(B_{j},A_{0})+(-1)^{j}C(B_{j},A_{1})}\\
\leq\sqrt{2}\sum_{j=0,1}\sqrt{1+(-1)^{j}\mathrm{Re}(\eta)},
\end{multline}
which completes the proof. Note that by swapping the roles of A and
B, a similar inequality is obtained where $\eta=C(B_{0},B_{1})$.
$\Square$

\textbf{Theorem \ref{Theorem 2}}. \emph{Proof}. The inequality \eqref{eq:ineq1}
implies 
\begin{multline}
(1-\abs{C(B_{j},A_{0})}^{2})(1-\abs{C(B_{j},A_{1})}^{2})\\
-\abs{\eta-C(B_{j},A_{0})^{\dagger}C(B_{j},A_{1})}^{2}\geq0,\label{eq:basic}
\end{multline}
which follows from the non-negativity of the determinant of the matrix
obtained by subtracting the right hand side from the left hand side
in \eqref{eq:ineq1}. Therefore, 
\begin{multline}
\abs{\eta-C(B_{j},A_{0})^{\dagger}C(B_{j},A_{1})}\\
\leq\sqrt{(1-\abs{C(B_{j},A_{0})}^{2})(1-\abs{C(B_{j},A_{1})}^{2})}.
\end{multline}
This and the triangle inequality give rise to the theorem, 
\begin{multline}
\abs{C(B_{0},A_{0})^{\dagger}C(B_{0},A_{1})-C(B_{1},A_{0})^{\dagger}C(B_{1},A_{1})}\\
\leq\sum_{j=0,1}\abs{\eta-C(B_{j},A_{0})^{\dagger}C(B_{j},A_{1})}\\
\leq\sum_{j=0,1}\sqrt{(1-\abs{C(B_{j},A_{0})}^{2})(1-\abs{C(B_{j},A_{1})}^{2})}.\quad\Square
\end{multline}

\textbf{Theorem \ref{Theorem 3}}. \emph{Proof}. We have seen that
\begin{equation}
\abs{\mathcal{B}}\leq\sqrt{2}\left(\sqrt{1+\mathrm{Re}(\eta)}+\sqrt{1-\mathrm{Re}(\eta)}\right).
\end{equation}
Therefore, 
\begin{equation}
\abs{\mathcal{B}}^{2}=\mathrm{Re}(\mathcal{B})^{2}+\mathrm{Im}(\mathcal{B})^{2}\leq4\left(1+\sqrt{1-\mathrm{Re}(\eta)^{2}}\right).
\end{equation}
Because, $\sqrt{1-a}\leq1-a/2$ for $a\in[0,1]$, it follows that
\begin{equation}
\mathrm{Re}(\mathcal{B})^{2}+\mathrm{Im}(\mathcal{B})^{2}\leq8-2\mathrm{Re}(\eta)^{2},
\end{equation}
from which the theorem follows. $\Square$

\textbf{Theorem \ref{Theorem 4}}. \emph{Proof}. In case the isotropy
holds, i.e., $C(A_{i},B_{j})=C(B_{j},A_{i})^{*}=(-1)^{ij}\varrho$,
\eqref{eq:basic} reads 
\begin{equation}
\abs{\eta-(-1)^{j}\abs{\varrho}^{2}}^{2}\leq(1-\abs{\varrho}^{2})^{2},
\end{equation}
and thus 
\begin{equation}
\abs{\eta}^{2}-2(-1)^{j}\abs{\varrho}^{2}\mathrm{Re}(\eta)\leq1-2\abs{\varrho}^{2}.
\end{equation}
Averaging both sides in this inequality over $j=0,1$, and rearranging
give 
\begin{equation}
\abs{\eta}^{2}+2\abs{\varrho}^{2}\leq1.\label{eq:circ}
\end{equation}
Finally, substituting $\varrho=\mathcal{B}/4$ into \eqref{eq:circ}
yields the theorem. $\Square$

\section{Investigating the bounds in the system of parafermions}

\label{sec:2}

Parafermions provide a unique test system for the bounds proven in
the previous section. First, the natural observables in a system of
parafermions are non-Hermitian. Second, in this system the non-commutativity
between Alice's and Bob's operators can be switched on and off without
changing anything else about the algebra of operators, enabling a
clean investigation of the effect of Alice-Bob non-commutativity.
Finally, there have been a number of proposals for experimental implementations
of parafermions \citep{Lindner2012,Cheng2012,Vaezi2013,Clarke2013,Mong2014,Klinovaja2014,Klinovaja2014a},
which opens the way for experimental verification of our predictions.

The structure of the section is as follows. In Sec.~\ref{sec:parafermion_physics+algebra+measurements},
we give a brief introduction to the physics of parafermions and the
algebra of their operators. In Sec.~\ref{sec:parafermions_Alice's-and-Bob's_observables},
we construct the observables of Alice and Bob. Those not interested
in the physics of parafermions may skip directly to Eqs.~(\ref{eq:observables_permutations_localA}--\ref{eq:observables_permutations_Alice-Bob})
detailing the permutation relations of the observables and Eqs.~(\ref{eq:observables_matrix_Alice0}--\ref{eq:observables_matrix_Bob'1})
introducing their explicit matrix representation. In Sec.~\ref{subsec:numerics},
we provide the results of the numerical investigation of bounds (\ref{Theorem 1 ineq}--\ref{Theorem 4 ineq}).

\subsection{\label{sec:parafermion_physics+algebra+measurements}Parafermion
physics and algebra}

\begin{figure}
\noindent \begin{centering}
\includegraphics[width=1\columnwidth]{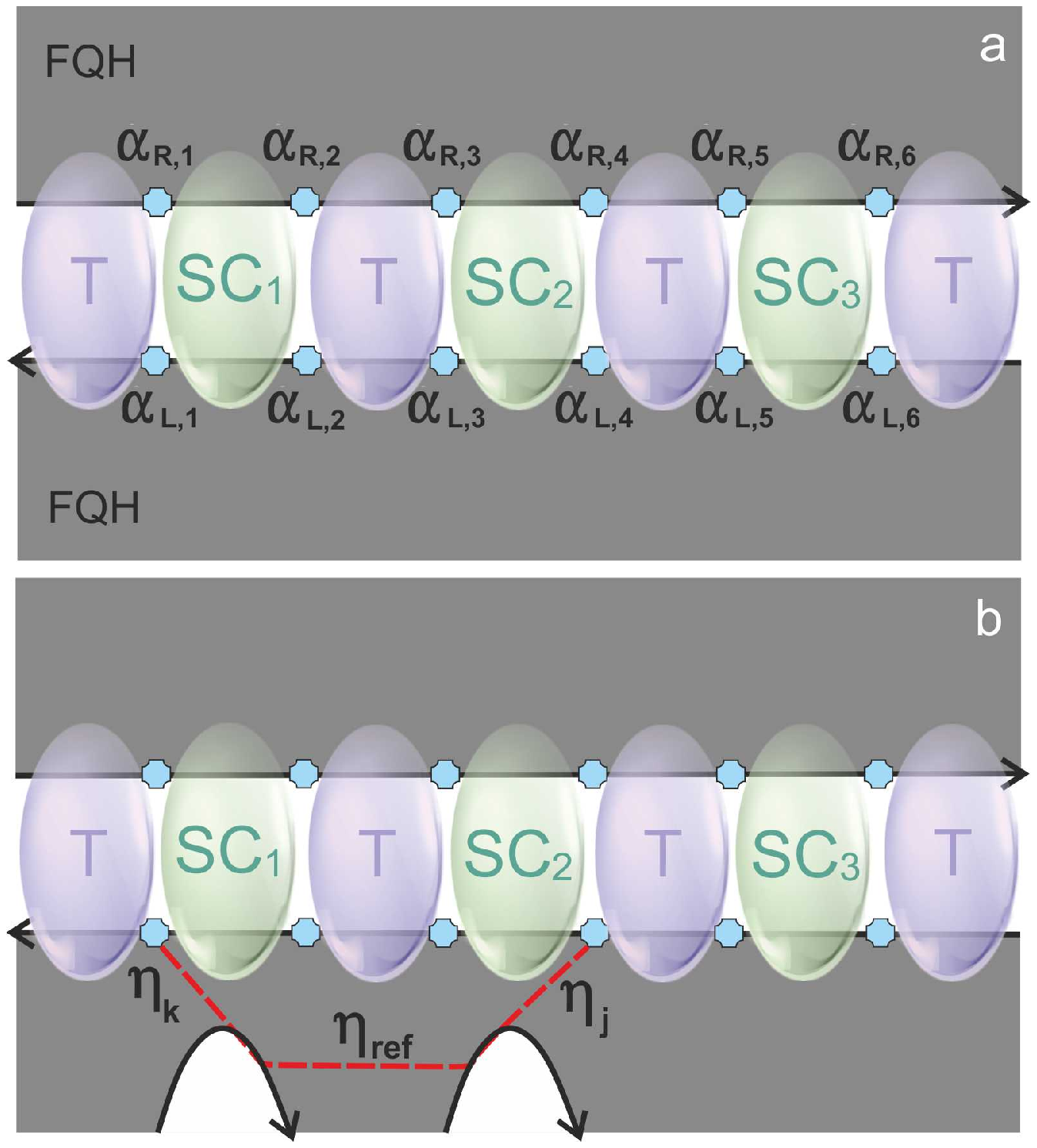} 
\par\end{centering}
\caption{\label{fig:parafermions_setup}A physical setup for creating and measuring
parafermions. a~--- Setup for implementing parafermions (represented
in cyan) with two fractional quantum Hall (FQH) edges (arrows) supporting
a series of electron-tunneling-gapped (T) and superconductivity-gapped
(SC) domains. b~--- Setup for measuring parafermionic observables
with the help of two additional FQH edges (curved arrows) as in Ref.~\citep{Snizhko2018}
(cf.~Appendix~\ref{subsec:parafermions_measurements}).}
\end{figure}

Parafermionic zero modes can be created in a variety of settings \citep{Lindner2012,Cheng2012,Vaezi2013,Clarke2013,Mong2014,Klinovaja2014,Klinovaja2014a}.
In different settings, they have subtly different properties. We focus
on parafermions implemented with the help of fractional quantum Hall
(FQH) edges proximitized by a superconductor \citep{Lindner2012,Clarke2013,Mong2014}.
The setup employs two FQH puddles of the same filling factor $\nu$
(grey regions in Fig.~\ref{fig:parafermions_setup}a) separated by
vacuum. This gives rise to two counter-propagating chiral FQH edges.
The edges can be gapped either by electron tunneling between them
(T domains) or by proximity-induced superconducting pairing of electrons
at the edges (SC domains). Domain walls between the domains of two
types host parafermionic zero modes $\alpha_{s,j}$ with $s=R/L=\pm1$
denoting whether a parafermion belongs to the right- or left-propagating
edge respectively, and $j$ denoting the domain wall number.

Parafermion operators have the following properties: 
\begin{equation}
\alpha_{s,j}^{2/\nu}=\alpha_{s,j}^{\,}\alpha_{s,j}^{\dagger}=\alpha_{s,j}^{\dagger}\alpha_{s,j}^{\,}=1,\label{eq:PF_properties_first}
\end{equation}
\begin{equation}
\alpha_{s,j}\alpha_{s,k}=\alpha_{s,k}\alpha_{s,j}e^{i\pi\nu s\sgn(k-j)},
\end{equation}
\begin{equation}
\alpha_{R,j}\alpha_{L,k}=\alpha_{L,k}\alpha_{R,j}\begin{cases}
e^{i\pi\nu} & ,\:k\neq j,\\
1 & ,\:k=j\text{ are even},\\
e^{2i\pi\nu} & ,\:k=j\text{ are odd},
\end{cases}\label{eq:PF_properties_last}
\end{equation}
where $\sgn$ is the sign function. These properties are valid for
$\nu=1/(2m+1)$, $m\in\mathbb{Z}_{+}$ considered in Refs.~\citep{Lindner2012,Clarke2013}
and for $\nu=2/3$ considered in Ref.~\citep{Mong2014}. In the case
of $\nu=1$, parafermions reduce to Majorana operators and $\alpha_{R,j}=\alpha_{L,j}$.

The physics of parafermions is associated with degenerate ground states
of the system. Namely, beyond hosting Cooper pairs, each superconducting
domain $\mathrm{SC}_{j}$ can host a certain charge $Q_{j}$ ($\mathrm{mod}\:2e$)
quantized in the units of charge of FQH quasiparticles $\nu e$. Thus
each $Q_{j}$ has $d=2/\nu$ distinct values, and the ground state
degeneracy of a system as in Fig.~\ref{fig:parafermions_setup}a
is therefore $d^{N_{\mathrm{SC}}}$, where $N_{\mathrm{SC}}$ is the
number of SC domains. Parafermionic operators $\alpha_{s,j}$ act
in this degenerate space of ground states and represent the effect
of adding a FQH quasiparticle to the system from a FQH puddle corresponding
to $s$ at domain wall $j$. Various observables in the system of
parafemions can be expressed through unitary operators $\alpha_{s,j}^{\,}\alpha_{s,k}^{\dagger}$.
In particular, $Q_{j}$ themselves can be expressed through $e^{i\pi s(Q_{j}/e-\nu/2)}=(-1)^{2/\nu}\alpha_{s,2j-1}^{\dagger}\alpha_{s,2j}^{\,}$.
One can show that $\left(\alpha_{s,j}^{\,}\alpha_{s,k}^{\dagger}\right)^{d}=-e^{2i\pi/\nu}$,
which implies that $\alpha_{s,j}^{\,}\alpha_{s,k}^{\dagger}$ has
$d$ distinct eigenvalues, all having the form $-e^{i\pi\nu(r+1/2)}$
with $r\in\mathbb{Z}$.

Unitary operators $\alpha_{s,j}^{\,}\alpha_{s,k}^{\dagger}$ are thus
natural ``observables'' in the system despite being non-Hermitian.
The permutation relations of such operators immediately follow from
Eqs.~(\ref{eq:PF_properties_first}-\ref{eq:PF_properties_last}).
Despite being spatially disconnected, such operators composed of different
pairs of parafermions may not commute, e.g., 
\begin{equation}
\alpha_{R,2}^{\,}\alpha_{R,4}^{\dagger}\alpha_{R,3}^{\,}\alpha_{R,5}^{\dagger}=\alpha_{R,3}^{\,}\alpha_{R,5}^{\dagger}\alpha_{R,2}^{\,}\alpha_{R,4}^{\dagger}e^{2i\pi\nu}.
\end{equation}
It is interesting to note that in the case of Majoranas ($\nu=1$),
none of these two unique properties would hold: the operators $i\alpha_{s,j}^{\,}\alpha_{s,k}^{\dagger}$
would be Hermitian, while two such operators having no common Majoranas
would commute.

\subsection{\label{sec:parafermions_Alice's-and-Bob's_observables}Alice's and
Bob's observables}

\begin{figure}
\begin{centering}
\includegraphics[width=1\columnwidth]{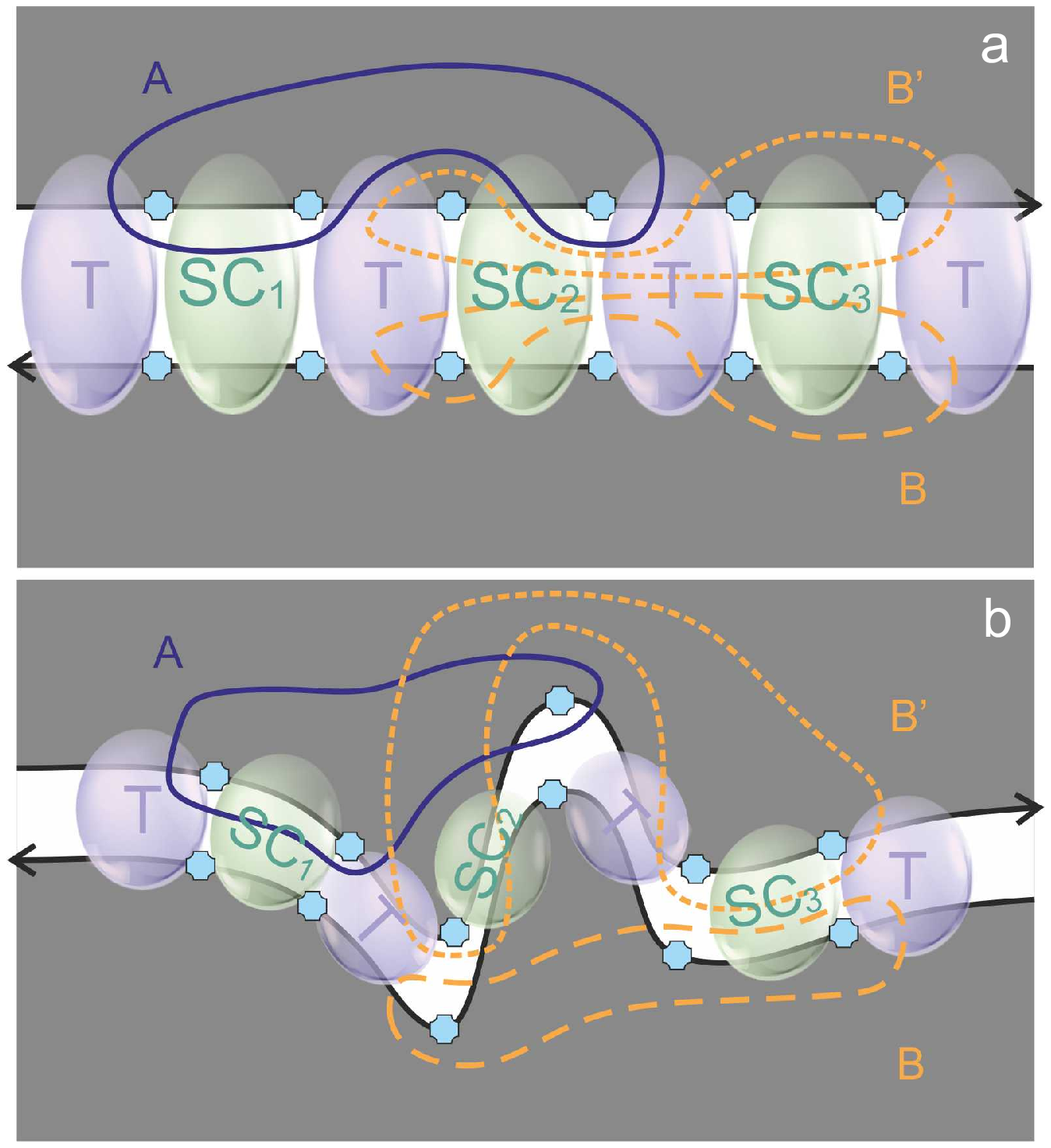} 
\par\end{centering}
\caption{\label{fig:PFs_AandB}Parafermionic observables and their mutual locality.
a~--- Grouping parafermions into groups belonging to Alice ($A$)
and Bob ($B$/$B'$). b~--- $A$ and $B$ do not have common parafermions,
are mutually local, and can be made arbitrarily distant in space.
While $A$ and $B'$ do not have common parafermions, they are not
mutually local: for Alice to measure $A$ while Bob can measure $B'$,
there should be a region of the upper FQH puddle accessible both to
Alice and Bob.}
\end{figure}

For a parfermionic system with three SC domains (as in Fig.~\ref{fig:parafermions_setup})
with a fixed total charge, the ground state is $d^{2}$-degenerate,
which allows to split it into two distinct subsystems: $\mathrm{SC}_{1}$
and $\mathrm{SC}_{3}$ domains, each having degeneracy $d$ as each
can have $d$ distinct values of charge $Q_{j}$. The charge of $\mathrm{SC}_{2}$
domain is determined by the state of $\mathrm{SC}_{1}$ and $\mathrm{SC}_{3}$
in order for the total charge to be fixed. This system is thus a natural
candidate for studying quantum correlations between two subsystems.
To this end, we introduce observables accessible to Alice, 
\begin{equation}
A_{0}=\alpha_{R,2}^{\,}\alpha_{R,4}^{\dagger},\quad A_{1}=\alpha_{R,1}^{\,}\alpha_{R,4}^{\dagger},\label{eq:observables_Alice}
\end{equation}
and two different sets of observables accessible to Bob: 
\begin{equation}
B_{0}=\alpha_{L,3}^{\,}\alpha_{L,5}^{\dagger},\quad B_{1}=\alpha_{L,3}^{\,}\alpha_{L,6}^{\dagger},\label{eq:observables_Bob}
\end{equation}
and 
\begin{equation}
B_{0}'=\alpha_{R,3}^{\,}\alpha_{R,6}^{\dagger},\quad B_{1}'=\alpha_{R,3}^{\,}\alpha_{R,5}^{\dagger}.\label{eq:observables_Bob'}
\end{equation}
They have identical local algebra, yet different commutation properties
of Alice's and Bob's observables:

\begin{equation}
A_{0}A_{1}=A_{1}A_{0}e^{-i\pi\nu},\label{eq:observables_permutations_localA}
\end{equation}

\begin{equation}
B_{0}B_{1}=B_{1}B_{0}e^{-i\pi\nu},\quad B_{0}'B_{1}'=B_{1}'B_{0}'e^{-i\pi\nu},\label{eq:observables_permutations_localB}
\end{equation}
\begin{equation}
\left[A_{j},B_{k}\right]=0,\quad A_{j}B_{k}'=B_{k}'A_{j}e^{2i\pi\nu}.\label{eq:observables_permutations_Alice-Bob}
\end{equation}
The non-commutation of $A$ and $B'$ observables would imply the
possiblity of superluminal signaling had the observables been truly
spatially separate (which is not the case, as we explain below). Therefore,
we call the set of $A$ and $B$ a non-signaling set, and the set
of $A$ and $B'$ a signaling set of observables.

Naively, Alice's observables are local with respect to either set
of Bob's observables, cf.~Fig.~\ref{fig:PFs_AandB}a. Indeed, $A$
and either the $B$ or $B'$ set use different parafermions, which
can be made arbitrarily distant from each other, cf.~Fig.~\ref{fig:PFs_AandB}b.
However, the locality issue in this system is subtler as in order
to probe an observable of the form $\alpha_{s,j}^{\,}\alpha_{s,k}^{\dagger}$,
one needs to enable FQH quasiparticle tunneling to both parafermions
simultaneously (see Appendix~\ref{subsec:parafermions_measurements}).
At the same time, quasiparticles can tunnel to a parafermion only
from the FQH puddle corresponding to the parafermion index $s$, not
through vacuum and not from the other puddle. Therefore, as can be
seen from Fig.~\ref{fig:PFs_AandB}b, the $A$ and $B$ sets are
indeed mutually local, while $A$ and $B'$ are not. The ability of
Alice to measure observables in $A$ and of Bob to measure observables
in $B'$, requires them to have access to a common region of the upper
FQH puddle. Thus, the system does not violate the laws of quantum
mechanics, nor exhibits superluminal signaling. Nevertheless, it presents
a unique opportunity for comparing correlations of commuting and non-commuting
(but otherwise equivalent) sets of observables.

The standard tool for studying quantum correlations is given by Bell
inequalities. However, since the observables considered here have
more than two eigenvalues, we require CHSH-like inequalities suitable
for multi-outcome measurements. We study the inequalities introduced
in Theorems~\ref{Theorem 1}--\ref{Theorem 4}, as well as an inequality
from Ref.~\citep{Fu2004}. These inequalities involve correlators
of the form $\langle A_{j}B_{k}^{\dagger}\rangle$ and $\langle A_{j}\left(B'_{k}\right)^{\dagger}\rangle$.
Since $\left[A_{j},B_{k}\right]=0$, $\langle A_{j}B_{k}^{\dagger}\rangle$
can be experimentally obtained by performing strong measurements of
$A_{j}$ and $B_{k}$ separately according to the protocol of Appendix~\ref{subsec:parafermions_measurements}
and then calculating the correlations. Alternatively, these correlations
can be measured with weak measurements \citep{Aharonov1988,Tamir2013}.
The non-commutativity of $A_{j}$ and $B_{k}'$ does not allow for
a strong-measurement-based approach in the case of $\langle A_{j}\left(B'_{k}\right)^{\dagger}\rangle$.
However, this correlator can be measured with the help of weak measurements
as described in Appendix~\ref{appendix:measuring_correlations_with_weak_measurements}.

From now on we focus on parafermions implemented using $\nu=2/3$
FQH puddles. Using permutation relations (\ref{eq:observables_permutations_localA}--\ref{eq:observables_permutations_Alice-Bob})
supplemented by the permutation relations of $B_{j}$ and $B_{k}'$,
as well as $\left(\alpha_{s,j}^{\,}\alpha_{s,k}^{\dagger}\right)^{3}=1$,
one can derive an explicit matrix representation for observables (\ref{eq:observables_Alice}--\ref{eq:observables_Bob'}):

\begin{gather}
A_{0}=\begin{pmatrix}1 & 0 & 0\\
0 & e^{2\pi i/3} & 0\\
0 & 0 & e^{-2\pi i/3}
\end{pmatrix}\otimes\begin{pmatrix}1 & 0 & 0\\
0 & 1 & 0\\
0 & 0 & 1
\end{pmatrix},\label{eq:observables_matrix_Alice0}\\
A_{1}=\begin{pmatrix}0 & 1 & 0\\
0 & 0 & 1\\
1 & 0 & 0
\end{pmatrix}\otimes\begin{pmatrix}1 & 0 & 0\\
0 & 1 & 0\\
0 & 0 & 1
\end{pmatrix},\label{eq:observables_matrix_Alice1}\\
B_{0}=\begin{pmatrix}1 & 0 & 0\\
0 & 1 & 0\\
0 & 0 & 1
\end{pmatrix}\otimes\begin{pmatrix}1 & 0 & 0\\
0 & e^{2\pi i/3} & 0\\
0 & 0 & e^{-2\pi i/3}
\end{pmatrix},\label{eq:observables_matrix_Bob0}\\
B_{1}=\begin{pmatrix}1 & 0 & 0\\
0 & 1 & 0\\
0 & 0 & 1
\end{pmatrix}\otimes\begin{pmatrix}0 & 1 & 0\\
0 & 0 & 1\\
1 & 0 & 0
\end{pmatrix},\label{eq:observables_matrix_Bob1}\\
B_{0}'=\begin{pmatrix}0 & e^{-2\pi i/3} & 0\\
0 & 0 & e^{2\pi i/3}\\
1 & 0 & 0
\end{pmatrix}\otimes\begin{pmatrix}0 & 1 & 0\\
0 & 0 & 1\\
1 & 0 & 0
\end{pmatrix},\label{eq:observables_matrix_Bob'0}\\
B_{1}'=\begin{pmatrix}0 & e^{-2\pi i/3} & 0\\
0 & 0 & e^{2\pi i/3}\\
1 & 0 & 0
\end{pmatrix}\otimes\begin{pmatrix}0 & 0 & 1\\
e^{-2\pi i/3} & 0 & 0\\
0 & e^{2\pi i/3} & 0
\end{pmatrix}.\label{eq:observables_matrix_Bob'1}
\end{gather}
This is used in the numerical investigation in the next section.

\subsection{\label{subsec:numerics}Numerical results for correlations of parafermions}

\begin{table*}
\begin{centering}
\begin{tabular}{|>{\raggedright}m{2.5cm}|>{\centering}m{1.8cm}|>{\centering}m{2cm}|>{\centering}m{3cm}|>{\centering}m{2cm}|>{\centering}m{2cm}|>{\centering}m{3cm}|}
\hline 
\multicolumn{2}{|c|}{\textbf{Bound:}} & \textbf{$I_{3}$ (\ref{eq:Li-bin_Fu_CHSH}),\,l.h.s.} & \textbf{Generalized Tsirelson (\ref{Theorem 1 ineq}),\,l.h.s.} & \textbf{Generalized TLM (\ref{Theorem 2 ineq}),}

\textbf{l.h.s/r.h.s.} & \textbf{Relation (\ref{Theorem 3 ineq}),\,l.h.s.} & \textbf{Relation (\ref{Theorem 4 ineq}),\,l.h.s.}\tabularnewline
\hline 
\multicolumn{2}{|c|}{Theoretical maximum} & $2.91$ \citep{Fu2003} & $2\sqrt{2}$

($\approx2.83$) & $1$ & $1$ & $1$ (if assumptions hold)\tabularnewline
\hline 
\multirow{2}{2.5cm}{Maximum for parafermionic observables} & non-signaling

($A$+$B$) & $2.60$ & $2.44$ & $0.71$ & $0.74$ & $1.00$\tabularnewline
\cline{2-7} \cline{3-7} \cline{4-7} \cline{5-7} \cline{6-7} \cline{7-7} 
 & signaling

($A$+$B'$) & $2.60$ & $2.82$ & $1.00$ & $1.00$ & $1.56$\tabularnewline
\hline 
\end{tabular}
\par\end{centering}
\caption{\label{tab:CHSH_results}Characterization of various bounds on non-local
correlations for the signaling and non-signaling sets of parafermionic
observables.}
\end{table*}

Here we numerically investigate the bounds on correlations presented
above (\ref{Theorem 1 ineq}--\ref{Theorem 4 ineq}) and the CHSH-like
inequality derived in Ref.~\citep{Fu2004}. The inequality of Ref.~\citep{Fu2004}
states that for a local-realistic system 
\begin{equation}
I_{3}=Q_{00}+Q_{01}-Q_{10}+Q_{11}\leq2,\label{eq:Li-bin_Fu_CHSH}
\end{equation}
where $Q_{jk}=\mathrm{Re}\left[\langle A_{j}B_{k}\rangle\right]+\frac{1}{\sqrt{3}}\mathrm{Im}\left[\langle A_{j}B_{k}\rangle\right]$
for $i\geq j$, and $Q_{01}=\mathrm{Re}\left[\langle A_{0}B_{1}\rangle\right]-\frac{1}{\sqrt{3}}\mathrm{Im}\left[\langle A_{0}B_{1}\rangle\right]$.
The observables are assumed to have possible values (for the quantum
case that we are interested in, eigenvalues) $e^{2\pi ir/3}$, $r\in\mathbb{Z}$,
which is the case for the observables defined in Eqs.~(\ref{eq:observables_matrix_Alice0}--\ref{eq:observables_matrix_Bob'1}).
In the standard quantum theory, i.e., for quantum observables such
that $\left[A_{j},B_{k}\right]=0$, the maximum attainable value is
known to be $\approx2.91$ \citep{Fu2003}.

For all the inequalities investigated, we calculated the corresponding
correlations $C(A_{i},B_{j})$ or $\langle A_{j}B_{k}\rangle$, and
maximized the relevant expressions numerically over all possible states
$\ket{\psi}$. The expressions maximized were the left-hand side of
bounds $(\ref{eq:Li-bin_Fu_CHSH},\ref{Theorem 1 ineq},\ref{Theorem 3 ineq},\ref{Theorem 4 ineq})$
and the ratio of the left-hand side to the right-hand side of inequality
(\ref{Theorem 2 ineq}). The numerical maximization was performed
independently via Wolfram Mathematica (functions \texttt{NMaximize}
for finding the global maximum and \texttt{FindMaximum} for investigating
local maxima) and Python (package \texttt{scipy.optimize}, functions
\texttt{basinhopping} for finding the global maximum with \texttt{SLSQP}
method for investigating local maxima). One aspect deserves mentioning.
Correlation functions $C(A_{i},B_{j})$ defined in Eq.~(\ref{eq:correlations_def})
are not well-defined in all of the Hilbert space as the denominator
can turn out to be zero. However, the points where it does, constitute
a set of measure zero among all the states. Moreover, in the vicinity
of these special points, $C(A_{i},B_{j})$ does not diverge but stays
bounded as $\abs{C(A_{i},B_{j})}\leq1$; however, the limiting value
as one approaches the special point depends on the direction of approach.
Therefore, with careful treatment, these special points do not constitute
a problem for investigation. Namely, we replaced $\cv{A_{i}}\rightarrow\cv{A_{i}}+\epsilon^{2}$,
$\cv{B_{j}}\rightarrow\cv{B_{j}}+\epsilon^{2}$ with a small cutoff
$\epsilon$, and checked that our results do not change as $\epsilon\rightarrow0$.
Furthermore, the states $\ket{\psi}$ on which the maximum values
in Table~\ref{tab:CHSH_results} are achieved are such that $\cv{A_{i}},\cv{B_{j}}\neq0$
for all $A_{i}$, $B_{j}$.

The results of our investigation are presented in Table~\ref{tab:CHSH_results}.
First, we note that the l.h.s. of Eq.~(\ref{eq:Li-bin_Fu_CHSH})
does not distinguish the signaling and non-signaling sets of observables.
Second, our bounds (\ref{Theorem 1 ineq}-\ref{Theorem 3 ineq}) are
obeyed by both sets. However, the signaling set saturates the bounds
much better than the non-signaling one. Finally, the bound of Theorem~\ref{Theorem 4},
(\ref{Theorem 4 ineq}), is saturated by the non-signaling set and
\emph{violated} by the signaling one. This does not contradict the
proof, which assumes $C(A_{i},B_{j})=(-1)^{ij}\varrho$. In fact,
this property is not satisfied by the states $\ket{\psi}$ maximizing
the l.h.s. of (\ref{Theorem 4 ineq}) for either of the sets. However,
this numerical evidence together with the fact that $C(A_{i},B_{j})=(-1)^{ij}\varrho$
correlations might be special for non-signaling theories (cf. the
discussion after Theorem~\ref{Theorem 4}) imply that Eq.~(\ref{Theorem 4 ineq})
may be a good bound for distinguishing signaling and non-signaling
quantum theories. We provide further evidence for the last statement
in Appendix~\ref{subsec:appendix_numerics}.

\section{Discussion}

\label{sec:3}

Our analytic results have important implications for understanding
quantum correlations. It is known that the standard CHSH parameter
has distinct bounds for classical local ($|\mathcal{B}|\leq2$) and
non-local ($|\mathcal{B}|\leq4$) hidden variable theories, while
the standard quantum theory obeys the Tsirelson bound (\ref{eq:Tsirelson_standard}).
Our variation of the Tsirelson bound (\ref{Theorem 1 ineq}) is closely
related to the original Tsirelson bound. In particular, for Hermitian
observables $X=A_{i},B_{j}$ such that $XX^{\dagger}=1$ and states
$\ket{\psi}$ such that $\bra{\psi}X\ket{\psi}=0$, our Bell-CHSH
parameter $|\mathcal{B}|$ (\ref{Theorem 1 ineq}) coincides with
the original one. At the same time, our proof shows that the Tsirelson
bound (\ref{Theorem 1 ineq}), as well as the TLM bound (\ref{Theorem 2 ineq}),
do not distinguish between the standard and non-local signaling quantum
theories. This implies that the Hilbert space structure is much more
restrictive than it was previously thought (see, e.g., Eq. \ref{semid}
which underlies our proofs). Naively, one could expect that the possibility
of signaling would allow nonlocal correlations to be stronger than
quantum, because one party can directly affect from a distance the
others' outcomes and in particular make them more correlated with
hers. However, the limited kind of signaling we have introduced here,
still within a quantum-like structure, is insufficient for this purpose.

At the same time, understanding the bounds on correlations in the
standard quantum theory, that explicitly takes into account the absence
of signaling, may be beneficial both for deepening its understanding,
further testing its validity, and deriving bounds on protocols for
quantum information processing. Our numerical results with parafermions
provide a candidate for such a bound, Eq.~(\ref{Theorem 4 ineq}).
Indeed, the ``non-signaling'' parafermionic set stayed within the
bound, while the ``signaling'' one violated it. Moreover, Ref.~\citep{Masanes2006}
argued that the assumptions we used to prove theorem \ref{Theorem 4}
hold generally for the states maximizing the standard Bell-CHSH parameter
in non-signaling theories (not in the sense that any maximizing state
satisfies the assumptions, but in the sense that it is always possible
to find a state that maximizes the standard Bell-CHSH parameter and
satisfies the assumptions). Therefore, we believe that inequality
(\ref{Theorem 4 ineq}) deserves further investigation.

\section*{Acknowledgements}

We wish to thank Yuval Gefen, Yuval Oreg, and Ivan S. Dotsenko for
helpful comments and discussions. Y.H. and E.C. wish to thank Yuval
Gefen, and also Ady Stern in the case of E.C., for hosting them during
September 2017 in the Weizmann Institute, where this work began. A.C.
acknowledges support from the Israel Science Foundation, Grant No.
1723/16. E.C. was supported by the Canada Research Chairs (CRC) Program.
K.S. would like to acknowledge funding by the Deutsche Forschungsgemeinschaft
(DFG, German Research Foundation) -- Projektnummer 277101999 --
TRR 183 (project C01) and the Italia-Israel project QUANTRA.

\appendix

\section*{Appendix}

\subsection{\label{sec:complex_probabilities}Relation between the correlation
functions $C(X,Y)$ and joint probability distributions}

For the standard case of commuting operators $X$ and $Y$, it is
possible to express correlators $C(X,Y)$ defined in Eq.~(\ref{eq:C(X,Y)_def})
through the joint probability distribution $P(x,y)$ of outcomes of
$X$ and $Y$ measurements. Indeed, for commuting $X$ and $Y$, it
is possible to find their common eigenbasis $\ket{xy\lambda}$, where
$X\ket{xy\lambda}=x\ket{xy\lambda}$ and similarly for $Y$; $\lambda$
represents possible additional quantum numbers. Then any state allows
for a decomposition 
\begin{equation}
\ket{\psi}=\sum_{x,y,\lambda}\alpha_{xy\lambda}\ket{xy\lambda}.
\end{equation}
The probability of one observer obtaining $x$ in a measurement of
$X$, while the other obtains $y$ in a measurement of $Y$ is given
by
\begin{equation}
P(x,y)=\mathrm{Tr}\ket{\psi}\bra{\psi}\mathcal{P}_{x}^{(X)}\mathcal{P}_{y}^{(Y)}=\sum_{\lambda}\abs{\alpha_{ab\lambda}}^{2},\label{eq:joint_prob}
\end{equation}
where $\mathcal{P}_{x}^{(X)}$ and $\mathcal{P}_{y}^{(Y)}$ are the
projectors onto the corresponding eigenspaces of $X$ and $Y$ respectively.
Then $\langle XY^{\dagger}\rangle=\sum_{x,y}xy^{*}P(x,y)$, $\langle X\rangle=\sum_{x}xP(x,y)$
etc. This allows for expressing $C(X,Y)$ as a non-linear functional
of the probability distribution $P(x,y)$. Therefore, for the case
of commuting Alice-Bob observables, $[A_{i},B_{j}]=0$ our bounds
(\ref{Theorem 1 ineq},\ref{Theorem 2 ineq}) can be considered restrictions
on the possible joint probability distributions $P(a,b|i,j)$ in the
quantum theory, defined exactly as in Eq.~(\ref{eq:joint_prob})
modulo a replacement $X\rightarrow A_{i}$ and $Y\rightarrow B_{j}$.

For the case of non-commuting $X$ and $Y$, one cannot define a joint
eigenbasis, but rather eigenbases $\ket{x\lambda}$ of $X$ and $\ket{y\tilde{\lambda}}$
of $Y$. One can still expand any state
\begin{equation}
\ket{\psi}=\sum_{x}\alpha_{x\lambda}\ket{x\lambda}
\end{equation}
and define
\begin{multline}
W(x,y)=\mathrm{Tr}\ket{\psi}\bra{\psi}\mathcal{P}_{x}^{(X)}\mathcal{P}_{y}^{(Y)}\\
=\sum_{x',\lambda,\lambda',\tilde{\lambda}}\alpha_{x'\lambda'}\alpha_{x\lambda}^{*}\sp{x\lambda}{y\tilde{\lambda}}\sp{y\tilde{\lambda}}{x'\lambda'}.\label{eq:joint_qprob}
\end{multline}
Moreover, $\langle XY^{\dagger}\rangle=\sum_{x,y}xy^{*}W(x,y)$, $\langle X\rangle=\sum_{x}xW(x,y)$
etc., leading to exactly the same expression of $C(X,Y)$ in terms
of $W(x,y)$ as previously in terms of $P(x,y)$. However, $W(x,y)$
is not a probability distribution as the r.h.s. of Eq.~(\ref{eq:joint_qprob})
can acquire complex values. $W(x,y)$ is a quasiprobability distribution
(somewhat similar to the Wigner function) in the case of non-commuting
$X$ and $Y$. Therefore, when $[A_{i},B_{j}]\neq0$ can be considered
as restrictions on the possible joint quasiprobability distributions
$W(a,b|i,j)$.

\subsection{\label{subsec:parafermions_measurements}Measuring parafermionic
observables}

A system combining parafermions with charging energy was introduced
in Ref.~\citep{Snizhko2018}. In such a system there is a charging
energy associated with the total system charge $Q_{\mathrm{tot}}=\sum_{j}Q_{j}+Q_{0}$,
where $Q_{0}=2en_{C}$ is the charge of the proximitizing superconductor,
and $n_{C}$ is the number of Cooper pairs in it. However, no energy
cost is associated with different distributions of a given total charge
over different SC domains. Therefore, the ground state of such a system
has degeneracy $d^{N_{\mathrm{SC}}-1}$, where the reduction by a
factor of $d$ corresponds to fixing the system's total charge. The
properties of operators $\alpha_{s,j}^{\,}\alpha_{s,k}^{\dagger}$
acting in this reduced subspace are identical to those in the original
system of parafermions with unrestricted total charge.

Introducing charging energy allows for designing a relatively simple
protocol for measuring $\alpha_{s,j}^{\,}\alpha_{s,k}^{\dagger}$
(both parafermions have the same $s$!) \citep{Snizhko2018}. A sketch
of the measurement setup is shown in Fig.~\ref{fig:parafermions_setup}b.
Two additional FQH edges (belonging to one of the puddles) are required
in this setup. Tunneling of FQH quasiparticles is allowed directly
between the two edges with tunnelling amplitude $\eta_{\mathrm{ref}}$
or between each edge and the corresponding parafermion $\alpha_{s,j/k}$
with amplitude $\eta_{j/k}$. As changing the charge of the parafermionic
system is energetically costly, the leading non-trivial process resulting
from coupling of the edges to the parafermions is co-tunneling of
quasiparticles: a quasiparticle is transferred between the edges,
while the parafermion state is changed via $\alpha_{s,j}^{\,}\alpha_{s,k}^{\dagger}$
and the effective tunneling amplitude is $\eta_{\mathrm{cot}}\simeq-\eta_{k}\eta_{j}^{*}/E_{C}$,
where $E_{C}$ is the charging energy. The two processes, direct and
parafermion-mediated tunneling of a quasiparticle between the edges,
interfere quantum-mechanically. When a voltage bias $V$ is applied
between the edges, the tunneling current between the edges is sensitive
to this interference: 
\begin{multline}
I_{\mathrm{T}}\propto\abs V^{2\nu-1}\sgn V\\
\times\left(\abs{\eta_{\mathrm{ref}}}^{2}+\abs{\eta_{\mathrm{cot}}}^{2}+2\kappa\mathrm{Re}\left[\eta_{\mathrm{ref}}^{*}\eta_{\mathrm{cot}}\alpha_{s,j}^{\,}\alpha_{s,k}^{\dagger}\right]\right),\label{eq:tunneling_current}
\end{multline}
where $\kappa$ is the interference suppression factor due to finite
temperature and other effects, $\mathrm{Re}\left[A\right]=\left(A+A^{\dagger}\right)/2$,
and $\abs V$ is assumed to be much larger than the temperature $T$
of the probing edges. As a result, by measuring $I_{\mathrm{T}}$,
one can measure the operator $\mathrm{Re}\left[e^{i\varphi}\alpha_{s,j}^{\,}\alpha_{s,k}^{\dagger}\right]$
with phase $\varphi$ depending on the phases of $\eta_{\mathrm{ref}}$
and $\eta_{\mathrm{cot}}$. Thus, one can measure the system in the
eigenstates of $\alpha_{s,j}^{\,}\alpha_{s,k}^{\dagger}$ employing
the fact that the eigenvalues of the $\alpha_{s,j}^{\,}\alpha_{s,k}^{\dagger}$
are discrete: for a generic $\varphi$, distinct eigenvalues of $\alpha_{s,j}^{\,}\alpha_{s,k}^{\dagger}$
correspond to distinct eigenvalues of $\mathrm{Re}\left[e^{i\varphi}\alpha_{s,j}^{\,}\alpha_{s,k}^{\dagger}\right]$.
Alternatively, through tuning the phase $\varphi$, one can measure
independently $\mathrm{Re}\left[\alpha_{s,j}^{\,}\alpha_{s,k}^{\dagger}\right]$
and $\mathrm{Im}\left[\alpha_{s,j}^{\,}\alpha_{s,k}^{\dagger}\right]=\mathrm{Re}\left[e^{-i\pi/2}\alpha_{s,j}^{\,}\alpha_{s,k}^{\dagger}\right]$,
and combine the measurement results for calculating the expectation
value $\langle\alpha_{s,j}^{\,}\alpha_{s,k}^{\dagger}\rangle$.

\subsection{\label{appendix:measuring_correlations_with_weak_measurements}How
to measure correlations of non-commuting observables}

\subsubsection{\textmd{\textup{\label{sec:measuring_non-commuting_parafermions}}}Measuring
correlations of non-commuting parafermionic observables}

Here we discuss how one can measure the correlators $\langle A_{j}\left(B'_{k}\right)^{\dagger}\rangle$
for non-commuting parafermionic observables. The procedure outlined
in Appendix~\ref{sec:measuring_non-commuting_observables} enables
one to measure $\langle\{A_{j},\left(B'_{k}\right)^{\dagger}\}\rangle$,
where $\{X,Y\}$ denotes the anti-commutator of operators $X$ and
$Y$, using weak measurements \citep{Aharonov1988,Tamir2013}. For
the observables defined in Sec.~\ref{sec:parafermions_Alice's-and-Bob's_observables},
the following permutation relation holds: $A_{j}\left(B'_{k}\right)^{\dagger}=\left(B'_{k}\right)^{\dagger}A_{j}e^{-2i\pi\nu}$.
Therefore, $\langle\{A_{j},\left(B'_{k}\right)^{\dagger}\}\rangle=\langle A_{j}\left(B'_{k}\right)^{\dagger}(1+e^{2i\pi\nu})\rangle=2\langle A_{j}\left(B'_{k}\right)^{\dagger}\rangle e^{i\pi\nu}\cos\pi\nu$,
and measuring $\langle\{A_{j},\left(B'_{k}\right)^{\dagger}\}\rangle$
is sufficient for measuring $\langle A_{j}\left(B'_{k}\right)^{\dagger}\rangle$.

The rest of this appendix is dedicated to designing weak measurements
of the required type and adapting the protocol of Appendix~\ref{sec:measuring_non-commuting_observables}
to measuring parafermionic observables. Note that this measurement
method is specific to the particular implementation of parafermions.
We start with the measurement protocol discussed in Appendix~\ref{subsec:parafermions_measurements}.
Suppose one of the additional FQH edges involved in the protocol has
voltage $V$ applied to it, while the other edge is grounded. The
current injected to the first edge is $I_{\mathrm{in}}=\nu e^{2}V/h$,
while the tunneling current between the edges is $I_{\mathrm{T}}$,
cf. Eq.~(\ref{eq:tunneling_current}). Suppose one measures the current
for time $t$, so that the number of quasiparticles injected into
the system is $N=I_{\mathrm{in}}t/(\nu e)$. The number of quasiparticles
$q$ tunneling within the time window will be fluctuating around the
average $\langle q\rangle=pN=I_{\mathrm{T}}t/(\nu e)$ with $p=I_{\mathrm{T}}/I_{\mathrm{in}}$.
The expression for $I_{\mathrm{T}}$ in Eq.~(\ref{eq:tunneling_current})
is valid as long as $\abs{I_{\mathrm{T}}}\ll\abs{I_{\mathrm{in}}}$.
In this regime, tunneling of different quasiparticles can be considered
independent, and thus the probability of observing tunneling of $q$
quasiparticles should be approximated well by the binomial distribution
\begin{equation}
P(q)=C_{N}^{q}p^{q}(1-p)^{N-q},\quad C_{N}^{q}=\frac{N!}{q!(N-q)!}.
\end{equation}
If one measures for a sufficiently long time, i.e., $N\gg1$, the
binomial distribution is well-approximated by the Gaussian distribution
\begin{equation}
P(q)\approx\frac{1}{\sqrt{2\pi Np(1-p)}}\exp\left(-\frac{(q-pN)^{2}}{2Np(1-p)}\right).\label{eq:tunnelling_gaussian}
\end{equation}

Depending on the eigenvalue $-e^{i\pi\nu(r+1/2)}$ of the measured
observable $\alpha_{s,j}^{\,}\alpha_{s,k}^{\dagger}$, the tunneling
probability $p=p_{0}+\delta p_{r}$, with 
\begin{equation}
p_{0}\propto\abs{\eta_{\mathrm{ref}}}^{2}+\abs{\eta_{\mathrm{cot}}}^{2},
\end{equation}
\begin{equation}
\delta p_{r}\propto-2\kappa\abs{\eta_{\mathrm{ref}}}\abs{\eta_{\mathrm{cot}}}\cos\left(\pi\nu r+\pi\nu/2+\varphi\right),
\end{equation}
where $\varphi=\arg(\eta_{\mathrm{ref}}^{*}\eta_{\mathrm{cot}})$,
cf.~Eq.~(\ref{eq:tunneling_current}). From now on we assume $\abs{\eta_{\mathrm{cot}}}\ll\abs{\eta_{\mathrm{ref}}}$,
$p_{0}\ll1$ and $p_{0}N\gg1$. Then the average number of tunneled
quasiparticles is $\langle q\rangle_{r}=p_{0}N+\delta p_{r}N$, while
the size of fluctuations in the measured values of $q$ is of the
order $\sigma=\sqrt{2Np(1-p)}=\sqrt{2Np_{0}}\left(1+O(\abs{\eta_{\mathrm{cot}}/\eta_{\mathrm{ref}}},p_{0})\right)$.
The parameter determining the distinguishability of different $r$,
and thus the measurement strength, is $\delta p_{r}N/\sigma\propto\abs{\frac{\eta_{\mathrm{cot}}}{\eta_{\mathrm{ref}}}}\sqrt{N}$.
For sufficiently large $\abs{\frac{\eta_{\mathrm{cot}}}{\eta_{\mathrm{ref}}}}\sqrt{N}$,
the scheme thus implements a strong measurement, while $\abs{\frac{\eta_{\mathrm{cot}}}{\eta_{\mathrm{ref}}}}\sqrt{N}\ll1$
implies a weak measurement.

Denoting the initial state of parafermions as $\sum_{r}\psi_{r}\ket r$
and using some approximations, one can derive the state of the system
after switching on the tunnel couplings for time $t$, 
\begin{equation}
\ket{\tilde{\Phi}}=\sum_{r,q,\lambda}f_{\lambda}(q,r)\psi_{r}\left(\frac{\eta_{\mathrm{ref}}-\eta_{\mathrm{cot}}e^{i\pi\nu(r+1/2)}}{\abs{\eta_{\mathrm{ref}}-\eta_{\mathrm{cot}}e^{i\pi\nu(r+1/2)}}}\right)^{q}\ket r\ket{q,\lambda},\label{eq:phi_tilde}
\end{equation}
where $\lambda$ represents additional quantum numbers of the edges.
It follows from Eq.~(\ref{eq:tunnelling_gaussian}) that $\sum_{\lambda}f_{\lambda}^{*}(q,r)f_{\lambda}(q,r)=\mathcal{N}^{2}\exp\left[-\frac{\left(q-\langle q\rangle_{r}\right)^{2}}{2Np_{0}}\right]\left[1+O\left(\abs{\frac{\eta_{\mathrm{cot}}}{\eta_{\mathrm{ref}}}},p_{0}\right)\right]$
with normalization factor $\mathcal{N}=\left(2\pi Np_{0}\right)^{-1/4}$.
Having not performed the calculation, we make a plausible assumption
that also 
\begin{multline}
\sum_{\lambda}f_{\lambda}^{*}(q,r)f_{\lambda}(q,r')=\mathcal{N}^{2}\\
\times\exp\left[-\left(q-\frac{\langle q\rangle_{r}+\langle q\rangle_{r'}}{2}\right)^{2}\times\frac{1}{2Np_{0}}-\frac{(\langle q\rangle_{r}-\langle q\rangle_{r'})^{2}}{8Np_{0}}\right]\\
\times\left[1+O\left(\abs{\frac{\eta_{\mathrm{cot}}}{\eta_{\mathrm{ref}}}},p_{0}\right)\right].
\end{multline}
Further assuming $\abs{\frac{\eta_{\mathrm{cot}}}{\eta_{\mathrm{ref}}}}p_{0}N\ll1$,
we can neglect $\eta_{\mathrm{cot}}e^{i\pi\nu(r+1/2)}$ in Eq.~(\ref{eq:phi_tilde})
and obtain that for our purposes one can replace $\ket{\tilde{\Phi}}$
with 
\begin{multline}
\ket{\Phi}=\mathcal{N}\sum_{r,q}\psi_{r}\exp\left[-\frac{\left(q-\langle q\rangle_{r}\right)^{2}}{4Np_{0}}\right]\\
\times\left[1+O\left(\abs{\frac{\eta_{\mathrm{cot}}}{\eta_{\mathrm{ref}}}}p_{0}N,\abs{\frac{\eta_{\mathrm{cot}}}{\eta_{\mathrm{ref}}}},p_{0}\right)\right]\ket r\ket q,
\end{multline}
which brings us to weak measurements of the type considered in Appendix~\ref{sec:measuring_non-commuting_observables}.

Consider now two weak measurements accessing $A_{j}$ and $\left(B'_{k}\right)^{\dagger}$
performed one after the other, with the number of quasiparticles tunneled
in each of the measurements being $q_{1}$ and $q_{2}$. Repeating
the calculation of Appendix~\ref{sec:measuring_non-commuting_observables},
we obtain 
\begin{multline}
\langle(q_{1}-p_{0}N)(q_{2}-p_{0}N)\rangle\propto\langle\mathrm{Re}\left[e^{i\varphi}A_{j}\right]\mathrm{Re}\left[e^{i\varphi'}\left(B'_{k}\right)^{\dagger}\right]\rangle\\
\times\left[1+O\left(\abs{\frac{\eta_{\mathrm{cot}}}{\eta_{\mathrm{ref}}}}p_{0}N,\abs{\frac{\eta_{\mathrm{cot}}}{\eta_{\mathrm{ref}}}},p_{0}\right)\right].
\end{multline}
Using Eq.~(\ref{eq:non-Herm_corr_Herm_corr_decomposition}), one
sees that by choosing different phases $\varphi$, $\varphi'$, one
can measure $\langle\{A_{j},\left(B'_{k}\right)^{\dagger}\}\rangle=2\langle A_{j}\left(B'_{k}\right)^{\dagger}\rangle e^{i\pi\nu}\cos\pi\nu$.

\subsubsection{\label{sec:measuring_non-commuting_observables}Measuring correlations
of non-commuting observables with weak measurements}

Here we discuss how to measure the averages $\langle\{A,B\}\rangle$
of non-Hermitian non-commuting $A$ and $B$, where $\{A,B\}=AB+BA$,
with the help of weak measurements. Our protocol uses essentially
the same measurement procedure as in Refs.~\citep{Arthurs1965,Johansen2008,Ochoa2018},
and is similar in spirit (yet has important differences) to Refs.~\citep{Resch2004,Lundeen2009}.
We note in passing that by more elaborate methods, one can measure
also the expectation value of a commutator \citep{Bulte2018}. However,
measuring the anti-commutator will suffice for our purposes. We first
discuss how to measure correlations of Hermitian non-commuting observables,
and then generalize the scheme to non-Hermitian observables.

Suppose one wants to measure the average $\langle\{A,B\}\rangle=\bra{\psi}\{A,B\}\ket{\psi}$,
where $A$ and $B$ are \emph{Hermitian} non-commuting operators,
and $\ket{\psi}$ is some quantum state. Introduce the eigenbases
of $A$ and $B$: $A\ket a=a\ket a$, $B\ket b=b\ket b$. Any system
state $\ket{\psi}$ can then be written as $\ket{\psi}=\sum_{a}\psi_{a}\ket a=\sum_{a,b}\psi_{a}\ket b\sp ba$
with some coefficients $\psi_{a}$. We assumed that there is no degeneracy
in the spectra of $A$ and $B$; generalization of the below consideration
for the case with degeneracy is straightforward.

Consider two detectors, $D_{1}$ and $D_{2}$ each having coordinate
$Q_{j}$ and momentum $P_{j}$ operators, $\left[P_{j},Q_{k}\right]=-i\delta_{jk}$,
with $j$ and $k$ having values $1$ and $2$. Prepare the system
and detectors in initial state 
\begin{equation}
\ket{\Phi_{in}}=\ket{\psi}\ket{D_{1,in}}\ket{D_{2,in}},
\end{equation}
\begin{equation}
\ket{D_{j,in}}=\mathcal{N}\int dq_{j}\exp\left(-\frac{q_{j}^{2}}{2\sigma^{2}}\right)\ket{q_{j}},
\end{equation}
where $\ket{q_{j}}$ is an eigenstate of $Q_{j}$ with eigenvalue
$q_{j}$, and $\mathcal{N}=\left(\pi\sigma^{2}\right)^{-1/4}$.

The Hamiltonian describing the system and the detectors is 
\begin{equation}
H(t)=\lambda_{1}(t)H_{1}+\lambda_{2}(t)H_{2},
\end{equation}
\begin{equation}
H_{1}=P_{1}A,\quad H_{2}=P_{2}B,
\end{equation}
where the coupling constants $\lambda_{j}(t)=0$ except for $\lambda_{1}(t)=g/T$
for $t\in(0;T)$ and $\lambda_{2}(t)=g/T$ for $t\in(T;2T)$. Then
after the system has interacted with the detectors, their state is
\begin{multline}
\ket{\Phi}=e^{-igH_{2}}e^{-igH_{1}}\ket{\Phi_{in}}\\
=\mathcal{N}^{2}\sum_{a,b}\int dq_{1}dq_{2}\psi_{a}\sp ba\ket b\ket{q_{1}}\ket{q_{2}}\\
\times\exp\left(-\frac{(q_{1}-ga)^{2}}{2\sigma^{2}}-\frac{(q_{2}-gb)^{2}}{2\sigma^{2}}\right).
\end{multline}
Measuring $Q_{1}$ and $Q_{2}$ of the detectors and calculating their
correlations then yields the desired quantity. Indeed, 
\begin{multline}
\bra{\Phi}Q_{1}Q_{2}\ket{\Phi}=\mathcal{N}^{4}\sum_{a,a',b}\psi_{a}^{*}\psi_{a'}\sp ab\sp b{a'}\\
\times\int dq_{2}q_{2}\exp\left(-\frac{(q_{2}-gb)^{2}}{\sigma^{2}}\right)\\
\times\int dq_{1}q_{1}\exp\left(-\frac{(q_{1}-g(a+a')/2)^{2}}{\sigma^{2}}-\frac{g^{2}(a-a')^{2}}{4\sigma^{2}}\right)\\
=\sum_{a,a',b}\alpha_{a}^{*}\alpha_{a'}\sp ab\sp b{a'}\frac{g^{2}}{2}b(a+a')\exp\left(-\frac{g^{2}(a-a')^{2}}{4\sigma^{2}}\right).
\end{multline}
Provided that $g\abs{a-a'}\ll2\sigma$ for all $a$, $a'$ (which
is the condition for weakness of the measurement), one obtains 
\begin{multline}
\bra{\Phi}Q_{1}Q_{2}\ket{\Phi}\\
=\frac{g^{2}}{2}\sum_{a,a',b}\psi_{a}^{*}\bra a\left(a\ket bb\bra b+\ket bb\bra ba'\right)\psi_{a'}\ket{a'}\\
=\frac{g^{2}}{2}\bra{\psi}\{A,B\}\ket{\psi}.
\end{multline}

Suppose now one wants to measure $\langle\{A,B\}\rangle=\bra{\psi}\{A,B\}\ket{\psi}$
for \emph{non-Hermitian} $A$ and $B$. Define the real and imaginary
part of each operator: $R_{A}=(A+A^{\dagger})$, $I_{A}=i(A^{\dagger}-A)/2$,
and similarly for $B$. It is easy to see that $\{A,B\}=\{R_{A},R_{B}\}-\{I_{A},I_{B}\}+i\{I_{A},R_{B}\}+i\{R_{A},I_{B}\}$.
Then 
\begin{multline}
\langle\{A,B\}\rangle=\langle\{R_{A},R_{B}\}\rangle-\langle\{I_{A},I_{B}\}\rangle\\
+i\langle\{I_{A},R_{B}\}\rangle+i\langle\{R_{A},I_{B}\}\rangle.\label{eq:non-Herm_corr_Herm_corr_decomposition}
\end{multline}
Each of the averages in the r.h.s. can be measured using the protocol
for Hermitian observables outlined above. Then combining them according
to Eq.~(\ref{eq:non-Herm_corr_Herm_corr_decomposition}) yields the
desired correlation of non-Hermitian non-commuting observables.

\subsection{\label{subsec:appendix_numerics}Extra numerical data on the bounds
for correlations in the system of parafermions}

In the main text, Table~\ref{tab:CHSH_results}, we provided the
results of testing the bounds on correlations for two sets of observables,
non-signaling ($A_{0}$, $A_{1}$, $B_{0}$, $B_{1}$) and signaling
($A_{0}$, $A_{1}$, $B_{0}'$, $B_{1}'$). Here, in Table~\ref{tab:CHSH_results_extra},
we present the results for several more sets of observables. Namely,
we checked what happens when the roles of Alice's operators $A_{0}$
and $A_{1}$ are exchanged, and similarly for Bob. Apart from that,
we also tested the sets involving $B_{2}=B_{0}^{\dagger}B_{1}=\alpha_{L,5}^{\,}\alpha_{L,6}^{\dagger}$
and $B_{2}'=B_{0}'^{\dagger}B_{1}'=\alpha_{R,6}^{\,}\alpha_{R,5}^{\dagger}$;
$\left[B_{2},A_{j}\right]=\left[B_{2}',A_{j}\right]=0$, with $A_{j}$,
$B_{j}$, $B_{j}'$ defined in Eqs.~(\ref{eq:observables_matrix_Alice0}--\ref{eq:observables_matrix_Bob'1}).
In all the sets we tested, all Alice's and Bob's operators commute
when Bob uses unprimed observables; some of Alice's operators do not
commute with some of the Bob's observables when Bob uses primed observables,
$B_{j}'$.

Note that all the sets we have tested obey all bounds except for relation
(\ref{Theorem 4 ineq}). The latter is obeyed by all the non-signaling
sets (when Bob uses $B_{j}$ observables) and violated by all the
signaling sets (when Bob uses $B_{j}'$ observables). This strengthens
the numerical evidence that relation (\ref{Theorem 4 ineq}) is a
good candidate for quantifying the effect of signaling on quantum
correlations.

In principle, the system of parafermions has many more possible sets
of observables. First, assigning different parafermions to Alice and
Bob, one can have different local algebras at Alice's and Bob's sites,
as well as different Alice-Bob commutation relations. We investigate
them in part by switching the order of $A_{0}$ and $A_{1}$ etc.
or replacing $B_{1}$ with $B_{2}$ in Table~\ref{tab:CHSH_results_extra}.
While this does not exhaust all the possibilities, the numerical results
we do have, indicate that our conclusions are likely to hold in the
cases we did not check. An even richer set of algebras can be accessed
by using operators beyond quadratic in parafermions, e.g., $\left(\alpha_{s,j}^{\,}\alpha_{s,k}^{\dagger}\right)^{n}$
or $\alpha_{s,j}^{2}\alpha_{s,k}^{\dagger}\alpha_{s,l}^{\dagger}$,
as well as arbitrary linear combinations of quadratic operators, e.g.,
$x\alpha_{s,j}\alpha_{s,k}^{\dagger}+y\alpha_{s,j}\alpha_{s,l}^{\dagger}$.
While investigating our bounds with these would be an interesting
non-trivial check, we believe that the more important task is understanding
and proving the role of Theorem~\ref{Theorem 4} and bound (\ref{Theorem 4 ineq})
in the general context.

\begin{table*}
\begin{centering}
\begin{tabular}{|>{\raggedright}m{2.5cm}|>{\centering}m{1.8cm}|>{\centering}m{1.8cm}|>{\centering}m{2cm}|>{\centering}m{2cm}|>{\centering}m{2cm}|>{\centering}m{2cm}|>{\centering}m{3cm}|}
\hline 
\multicolumn{3}{|c|}{\textbf{Bound:}} & \textbf{$I_{3}$ (\ref{eq:Li-bin_Fu_CHSH}),\,l.h.s.}  & \textbf{Generalized Tsirelson (\ref{Theorem 1 ineq}),\,l.h.s.}  & \textbf{Generalized TLM (\ref{Theorem 2 ineq}),}

\textbf{l.h.s/r.h.s.}  & \textbf{Relation (\ref{Theorem 3 ineq}),\,l.h.s.}  & \textbf{Relation (\ref{Theorem 4 ineq}),\,l.h.s.}\tabularnewline
\hline 
\multicolumn{3}{|c|}{\textbf{Theoretical maximum}} & \textbf{$2.91$ \citep{Fu2003}}  & \textbf{$2\sqrt{2}$}

\textbf{($\approx2.83$)}  & \textbf{$1$}  & \textbf{$1$}  & \textbf{1 (if assumptions hold)}\tabularnewline
\hline 
\multirow{11}{2.5cm}{\textbf{Maximum for parafermionic observables}} & \textbf{Alice's operators}  & \textbf{Bob's operators}  &  &  &  &  & \tabularnewline
\cline{2-8} \cline{3-8} \cline{4-8} \cline{5-8} \cline{6-8} \cline{7-8} \cline{8-8} 
 & \textbf{$A_{0}$, $A_{1}$}  & \textbf{$B_{0}$, $B_{1}$}  & \textbf{$2.60$}  & \textbf{$2.44$}  & \textbf{$0.71$}  & \textbf{$0.74$}  & \textbf{$1.00$}\tabularnewline
\cline{2-8} \cline{3-8} \cline{4-8} \cline{5-8} \cline{6-8} \cline{7-8} \cline{8-8} 
 & \textbf{$A_{0}$, $A_{1}$}  & \textbf{$B_{0}'$, $B_{1}'$}  & \textbf{$2.60$}  & \textbf{$2.82$}  & \textbf{$1.00$}  & \textbf{$1.00$}  & \textbf{$1.56$}\tabularnewline
\cline{2-8} \cline{3-8} \cline{4-8} \cline{5-8} \cline{6-8} \cline{7-8} \cline{8-8} 
 & \textbf{$A_{1}$, $A_{0}$}  & \textbf{$B_{1}$, $B_{0}$}  & \textbf{$2.60$}  & \textbf{$2.44$}  & \textbf{$0.71$}  & \textbf{$0.74$}  & \textbf{$1.00$}\tabularnewline
\cline{2-8} \cline{3-8} \cline{4-8} \cline{5-8} \cline{6-8} \cline{7-8} \cline{8-8} 
 & \textbf{$A_{1}$, $A_{0}$}  & \textbf{$B_{1}'$, $B_{0}'$}  & \textbf{$2.60$}  & \textbf{$2.82$}  & \textbf{$1.00$}  & \textbf{$1.00$}  & \textbf{$1.56$}\tabularnewline
\cline{2-8} \cline{3-8} \cline{4-8} \cline{5-8} \cline{6-8} \cline{7-8} \cline{8-8} 
 & \textbf{$A_{0}$, $A_{1}$}  & \textbf{$B_{1}$, $B_{0}$}  & \textbf{$2.60$}  & \textbf{$2.22$}  & \textbf{$0.71$}  & \textbf{$0.62$}  & \textbf{$1.00$}\tabularnewline
\cline{2-8} \cline{3-8} \cline{4-8} \cline{5-8} \cline{6-8} \cline{7-8} \cline{8-8} 
 & \textbf{$A_{0}$, $A_{1}$}  & \textbf{$B_{1}'$, $B_{0}'$}  & \textbf{$2.60$}  & \textbf{$2.71$}  & \textbf{$1.00$}  & \textbf{$0.97$}  & \textbf{$1.56$}\tabularnewline
\cline{2-8} \cline{3-8} \cline{4-8} \cline{5-8} \cline{6-8} \cline{7-8} \cline{8-8} 
 & \textbf{$A_{0}$, $A_{1}$}  & \textbf{$B_{0}$, $B_{2}$}  & \textbf{$2.60$}  & \textbf{$2.44$}  & \textbf{$0.71$}  & \textbf{$0.74$}  & \textbf{$1.00$}\tabularnewline
\cline{2-8} \cline{3-8} \cline{4-8} \cline{5-8} \cline{6-8} \cline{7-8} \cline{8-8} 
 & \textbf{$A_{0}$, $A_{1}$}  & \textbf{$B_{0}'$, $B_{2}'$}  & \textbf{$2.00$}  & \textbf{$2.23$}  & \textbf{$1.00$}  & \textbf{$0.75$}  & \textbf{$1.50$}\tabularnewline
\cline{2-8} \cline{3-8} \cline{4-8} \cline{5-8} \cline{6-8} \cline{7-8} \cline{8-8} 
 & \textbf{$A_{0}$, $A_{1}$}  & \textbf{$B_{2}$, $B_{0}$}  & \textbf{$2.60$}  & \textbf{$2.22$}  & \textbf{$0.71$}  & \textbf{$0.62$}  & \textbf{$1.00$}\tabularnewline
\cline{2-8} \cline{3-8} \cline{4-8} \cline{5-8} \cline{6-8} \cline{7-8} \cline{8-8} 
 & \textbf{$A_{0}$, $A_{1}$}  & \textbf{$B_{2}'$, $B_{0}'$}  & \textbf{$2.00$}  & \textbf{$2.44$}  & \textbf{$1.00$}  & \textbf{$0.75$}  & \textbf{$1.50$}\tabularnewline
\hline 
\end{tabular}
\par\end{centering}
\caption{\label{tab:CHSH_results_extra}Characterization of bounds on non-local
correlations for various sets of parafermionic observables.}
\end{table*}

\bibliography{library}

\end{document}